\documentclass[aps,prd,twocolumn,superscriptaddress]{revtex4}
\usepackage{epsfig,epsf}
\usepackage{amsmath}
\usepackage{amsthm}
\usepackage{amsfonts}
\usepackage{amssymb}
\usepackage{dsfont}
\usepackage{multirow}
\usepackage{appendix}
\usepackage{slashed}
\usepackage[active]{srcltx}
\usepackage{psfrag}

\begin{document}

\title{Resonance $X(7300)$: excited $2S$ tetraquark or hadronic molecule $%
\chi_{c1}\chi_{c1}$?}
\date{\today}
\author{S.~S.~Agaev}
\affiliation{Institute for Physical Problems, Baku State University, Az--1148 Baku,
Azerbaijan}
\author{K.~Azizi}
\affiliation{Department of Physics, University of Tehran, North Karegar Avenue, Tehran
14395-547, Iran}
\affiliation{Department of Physics, Do\v{g}u\c{s} University, Dudullu-\"{U}mraniye, 34775
Istanbul, T\"{u}rkiye}
\author{B.~Barsbay}
\affiliation{Division of Optometry, School of Medical Services and Techniques, Do\v{g}u%
\c{s} University, 34775 Istanbul, T\"{u}rkiye}
\author{H.~Sundu}
\affiliation{Department of Physics Engineering, Istanbul Medeniyet University, 34700
Istanbul, T\"{u}rkiye}

\begin{abstract}
We explore the first radial excitation $X_{\mathrm{4c}}^{\ast}$ of the fully
charmed diquark-antidiquark state $X_{\mathrm{4c}}=cc\overline{c}\overline{c}
$ built of axial-vector components, and the hadronic molecule $\mathcal{M}%
=\chi_{c1}\chi_{c1}$. The masses and current couplings of these scalar
states are calculated in the context of the QCD two-point sum rule approach.
The full widths of $X_{\mathrm{4c}}^{\ast}$ and $\mathcal{M}$ are evaluated
by taking into account their kinematically allowed decay channels. We find
partial widths of these processes using the strong couplings $g_i^{\ast}$
and $G_i^{(\ast)}$ at the $X_{\mathrm{4c}}^{\ast}$($\mathcal{M}$%
)-conventional mesons vertices computed by means of the QCD three-point sum
rule method. The predictions obtained for the parameters $m=(7235 \pm 75)~%
\mathrm{MeV}$, $\Gamma=(144 \pm 18)~\mathrm{MeV}$ and $\widetilde{m}=(7180
\pm 120)~\mathrm{MeV}$, $\widetilde{\Gamma}=(169 \pm 21)~\mathrm{MeV}$ of
these structures, are compared with the experimental data of the CMS and
ATLAS Collaborations. In accordance to these results, within existing errors
of measurements and uncertainties of the theoretical calculations, both the
excited tetraquark and hadronic molecule may be considered as candidates to
the resonance $X(7300)$. Detailed analysis, however, demonstrates that the 
preferable model for $X(7300)$ is an admixture of the molecule $\mathcal{M}$ and 
sizeable part of $X_{\mathrm{4c}}^{\ast}$.
\end{abstract}

\maketitle


\section{Introduction}

\label{sec:Int} 

The multiquark hadrons composed of exclusively heavy quarks were in agenda
of researches from first years of the parton model and QCD. During past
decades much was done to investigate features of such particles, calculate
their parameters in the context of different models, study production and
decay mechanisms of these hadrons. Reports of the LHCb, ATLAS and CMS
Collaborations on $X$ resonances in the $6.2$-$7.3~\mathrm{GeV}$ mass range
became one of important experimental achievements in the physics of fully
charmed four-quark mesons \cite%
{LHCb:2020bwg,Bouhova-Thacker:2022vnt,CMS:2023owd}. The structures $X(6200)$%
, $X(6600)$, $X(6900)$ and $X(7300)$ observed by these experiments in the di-%
$J/\psi $ and $J/\psi \psi ^{\prime }$ mass distributions provide useful
information and allow one to compare numerous theoretical predictions with
the masses and widths of these states.

These discoveries generated new theoretical activities to explain observed
states, reveal their internal structures \cite%
{Zhang:2020xtb,Albuquerque:2020hio,Wang:2022xja,Dong:2022sef,Faustov:2022mvs,Becchi:2020mjz,Becchi:2020uvq,Dong:2020nwy, Dong:2021lkh,Liang:2021fzr,Niu:2022vqp,Yu:2022lak,Kuang:2023vac}%
. The fully heavy $X$ resonances were considered as scalar four-quark mesons
with diquark-antidiquark or hadronic molecule organizations \cite%
{Zhang:2020xtb,Albuquerque:2020hio,Wang:2022xja,Dong:2022sef,Faustov:2022mvs}%
. For example, the resonance $X(6900)$ may be a diquark-antidiquark state
with pseudoscalar ingredients, or hadronic molecule $\chi _{c0}\chi _{c0}$
\cite{Albuquerque:2020hio}. The structure $X(6200)$ was interpreted as a
ground-level tetraquark with the spin-parities $J^{\mathrm{PC}}=0^{++}$ or $%
1^{+-}$, whereas $X(6600)-$ as its first radial excitation \cite%
{Wang:2022xja}. The four structures $X(6200)-X(7300)$ were assigned to be
different excited tetraquark states \cite{Dong:2022sef,Faustov:2022mvs}.

Alternative scenarios explain appearance of the $X$ resonances by
coupled-channel effects. Thus, using this approach the authors of Ref. \cite%
{Dong:2020nwy} predicted existence of the near-threshold state $X(6200)$
with $J^{\mathrm{PC}}=0^{++}$ or $2^{++}$ in the di-$J/\psi $ system.
Coupled-channel effects may also generate a pole structure identified in
Ref.\ \cite{Liang:2021fzr} with $X(6900)$, and lead to emergence of a bound
state $X(6200)$, and resonances $X(6680)$ and $X(7200)$, which can be
classified as broad and narrow structures, respectively.

Production mechanisms of fully heavy tetraquarks in different processes
became topics for interesting investigations \cite%
{Feng:2023agq,Abreu:2023wwg}. Thus, inclusive production of fully charmed $S$%
-wave four-quark mesons at the LHC energies was studied in the
nonrelativistic QCD factorization framework in Ref.\ \cite{Feng:2023agq}.
Production of fully-heavy tetraquark states in $pp$ and $pA$ collisions
through the double parton scattering mechanism was considered in Ref.\ \cite%
{Abreu:2023wwg}, in which it was shown that a search for such states is
feasible in the future runs of LHC and in Future Circular Collider.

The fully heavy four-quark mesons were studied also in our articles \cite%
{Agaev:2023wua,Agaev:2023gaq,Agaev:2023ruu}. The scalar tetraquarks $X_{%
\mathrm{4c}}=cc\overline{c}\overline{c}$ and $X_{\mathrm{4b}}=bb\overline{b}%
\overline{b}$ built of axial-vector diquarks were explored in Ref.\ \cite%
{Agaev:2023wua}. It was demonstrated that $X_{\mathrm{4c}}$ with the mass $%
(6570\pm 55)~\mathrm{MeV}$ and full width $(110\pm 21)~\mathrm{MeV}$ is nice
candidate to the resonance $X(6600)$. The fully beauty state $X_{\mathrm{4b}}
$ has the mass $(18540\pm 50)~\mathrm{MeV}$ that is smaller than the $\eta
_{b}\eta _{b}$ threshold therefore it cannot be seen in $\eta _{b}\eta _{b}$
or $\Upsilon (1S)\Upsilon (1S)$ mass distributions. The $X_{\mathrm{4b}}$
can decay to open-beauty mesons through $\overline{b}b$ annihilation to
gluon(s) that triggers $X_{\mathrm{4b}}\rightarrow B^{+}B^{-}$ and other
decays \cite{Becchi:2020mjz}. Electromagnetic decays to photons and leptons
are alternative channels for transformation of $X_{\mathrm{4b}}$ to
conventional particles.

The scalar tetraquarks $T_{\mathrm{4c}}$ and $T_{\mathrm{4b}}$ composed of
pseudoscalar diquarks were explored in Ref.\ \cite{Agaev:2023gaq}, in which
we computed their masses and widths. The parameters $m=(6928\pm 50)~\mathrm{%
MeV}$ and $\widetilde{\Gamma }_{\mathrm{4c}}=(128\pm 22)~\mathrm{MeV}$ of $%
T_{\mathrm{4c}}$ are in excellent agreements with relevant CMS data,
therefore we interpreted it as the resonance $X(6900)$. The exotic meson $T_{%
\mathrm{4b}}$ decays to $\eta _{b}\eta _{b}$ pairs and can be detected in
the mass distribution of these mesons. It is interesting that the hadronic
molecule $\chi _{c0}\chi _{c0}$ (a brief form of $\chi _{c0}(1P)\chi
_{c0}(1P)$) has similar parameters and is another candidate to $X(6900)$
\cite{Agaev:2023ruu}. Hence, $X(6900)$ may be considered as a linear
superposition of the molecule $\chi _{c0}\chi _{c0}$ and diquark-antidiquark
state $T_{\mathrm{4c}}$.

The lowest lying structure among $X$ states is the resonance $X(6200)$, that
may be interpreted as the molecule $\eta _{c}\eta _{c}$. In fact, the mass $%
(6264\pm 50)~\mathrm{MeV}$ and full width $(320\pm 72)~\mathrm{MeV}$ of the
molecule $\eta _{c}\eta _{c}$ agree with the LHCb-ATLAS-CMS data \cite%
{Agaev:2023ruu}.

The last position in the list of new $X$ structures is held by the resonance
$X(7300)$. This state was detected in both the di-$J/\psi $ and $J/\psi \psi
^{\prime }$ mass distributions. In Ref.\ \cite{Agaev:2023wua}, we used this
fact to make assumptions about its nature, and argued that $X(7300)$ maybe
is the $2S$ radial excitation of the exotic meson $X(6600)$. Another option
for $X(7300)$ is the hadronic molecule model $\chi _{c1}(1P)\chi _{c1}(1P)$
(in what follows $\chi _{c1}\chi _{c1}$) that may have close parameters.

In the present article, we address problems connected with the resonance $%
X(7300)$ in attempts to describe its parameters in the four-quark model. To
this end, we calculate the mass and width of the first radial excitation $X_{%
\mathrm{4c}}^{\ast }$ of the diquark-antidiquark state $X_{\mathrm{4c}}$.
The full width of $X_{\mathrm{4c}}^{\ast }$ is evaluated using its
kinematically allowed decays to $J/\psi J/\psi $, $J/\psi \psi ^{\prime }$, $%
\eta _{c}\eta _{c}$, $\eta _{c}\eta _{c}(2S)$, $\eta _{c}\chi _{c1}$, $\chi
_{c0}\chi _{c0}$, and $\chi _{c1}\chi _{c1}$ mesons. We are also going to
perform the similar analysis in the case of the molecule $\mathcal{M=}\chi
_{c1}\chi _{c1}$. We will compare predictions for parameters of $X_{\mathrm{%
4c}}^{\ast }$ and $\mathcal{M}$ with experimental data, and each other to
make decision about the nature of $X(7300)$.

This article is organized in the following form: In Sec. \ref{sec:Excitation}%
, we explore the excited tetraquark $X_{\mathrm{4c}}^{\ast }$ and compute
its mass and full width. The same analysis for the molecule $\mathcal{M}$ is
carried out in Sec.\ \ref{sec:Molecule}. In the last Section \ref{sec:Disc},
we present our brief conclusions. Appendix contains the heavy quark
propagator and some of correlation functions used in the present analysis.


\section{Radially excited state $X_{\mathrm{4c}}^{\ast}$}

\label{sec:Excitation}


In this section, we explore the first radial excitation $X_{\mathrm{4c}%
}^{\ast }$ of the scalar tetraquark $X_{\mathrm{4c}}$ built of axial-vector
diquarks. The mass and current coupling of this state are computed by means
of the QCD\ two-point sum rule (SR) approach \cite%
{Shifman:1978bx,Shifman:1978by}. To evaluate partial widths of the
kinematically allowed decay channels of $X_{\mathrm{4c}}^{\ast }$, we are
going to employ the three-point sum rule method, which is necessary to find
strong couplings at corresponding three-particle vertices. It is worth
noting that the SR methods are powerful nonperturbative tools to study
conventional hadrons, but they can also be applied for analyses of
multiquark particles \cite{Nielsen:2009uh,Albuquerque:2018jkn,Agaev:2020zad}.


\subsection{Mass $m$ and coupling $f$ of $X_{\mathrm{4c}}^{\ast }$}


The sum rules for the mass $m$ and current coupling $f$ of the tetraquark $%
X_{\mathrm{4c}}^{\ast }$ can be extracted from analysis of the correlation
function
\begin{equation}
\Pi (p)=i\int d^{4}xe^{ipx}\langle 0|\mathcal{T}\{J(x)J^{\dag
}(0)\}|0\rangle ,  \label{eq:CorrF1}
\end{equation}%
where $\mathcal{T}$ is the time-ordered product of two currents, and $J(x)$
is the interpolating current for the states $X_{\mathrm{4c}}$ and $X_{%
\mathrm{4c}}^{\ast }$.

We model $X_{\mathrm{4c}}$ and $X_{\mathrm{4c}}^{\ast }$ as tetraquarks
built of the axial-vector diquark $c^{T}C\gamma _{\mu }c$ and axial-vector
antidiquark $\overline{c}\gamma _{\mu }C\overline{c}^{T}$. Then, the
interpolating current is determined by the expression
\begin{equation}
J(x)=c_{a}^{T}(x)C\gamma _{\mu }c_{b}(x)\overline{c}_{a}(x)\gamma ^{\mu }C%
\overline{c}_{b}^{T}(x),  \label{eq:Current1}
\end{equation}%
with $a$ and $b$ being color indices. In Eq.\ (\ref{eq:Current1}) $c(x)$ is $%
c$-quark fields, and $C$ is the charge conjugation matrix. The current $J(x)$
describes the diquark-antidiquark states with spin-parities $J^{\mathrm{PC}%
}=0^{++}$.

The ground-level particle with this quark content and quantum numbers is the
tetraquark $X_{\mathrm{4c}}$ which was investigated in our paper \cite%
{Agaev:2023wua}. We computed its mass $m_{0}$ and coupling $f_{0}$ by
employing the two-point SR approach. We took into account explicitly only
the ground-state term and included all other contributions to a class of
"higher resonances and continuum states". We refer to this standard
treatment as "ground-state+continuum" approximation.

To derive sum rules for $m$ and $f$, we express the correlation function $%
\Pi (p)$ in terms of $X_{\mathrm{4c}}$ and $X_{\mathrm{4c}}^{\ast }$
tetraquarks' masses and couplings. Having inserted a complete set of
intermediate states with the same content and quantum numbers of these
tetraquarks, and carried out integration over $x$, we get
\begin{eqnarray}
&&\Pi ^{\mathrm{Phys}}(p)=\frac{\langle 0|J|X_{\mathrm{4c}}(p)\rangle
\langle X_{\mathrm{4c}}(p)|J^{\dagger }|0\rangle }{m_{0}^{2}-p^{2}}  \notag
\\
&&+\frac{\langle 0|J|X_{\mathrm{4c}}^{\ast }(p)\rangle \langle X_{\mathrm{4c}%
}^{\ast }(p)|J^{\dagger }|0\rangle }{m^{2}-p^{2}}\cdots .  \label{eq:CorrF2}
\end{eqnarray}%
This expression contains two terms corresponding to the ground-state
particle $X_{\mathrm{4c}}$ with the mass $m_{0}$ and a contribution coming
from the first radially excited state, i.e., from $2S$ level tetraquark $X_{%
\mathrm{4c}}^{\ast }$. Here, the ellipses stand for the effects of higher
resonances and continuum states. This approach is "ground-level+first
excited state +continuum" approximation.

The $\Pi ^{\mathrm{Phys}}(p)$ can be simplified using the matrix elements
\begin{equation}
\langle 0|J|X_{\mathrm{4c}}(p)\rangle =f_{0}m_{0},\ \langle 0|J|X_{\mathrm{4c%
}}^{\ast }(p)\rangle =fm,  \label{eq:MatEl1}
\end{equation}%
where $f_{0}$ and $f$ are current couplings of the $X_{\mathrm{4c}}$ and $X_{%
\mathrm{4c}}^{\ast }$, respectively. Then, we get
\begin{equation}
\Pi ^{\mathrm{Phys}}(p)=\frac{f_{0}^{2}m_{0}^{2}}{m_{0}^{2}-p^{2}}+\frac{%
f^{2}m^{2}}{m^{2}-p^{2}}+\cdots .  \label{eq:CorrF3}
\end{equation}%
This function contains only the Lorentz structure proportional to $\mathrm{I}
$, hence the invariant amplitude $\Pi ^{\mathrm{Phys}}(p^{2})$ necessary for
our analysis is defined by rhs of Eq.\ (\ref{eq:CorrF3}).

The QCD side of the sum rules is formed by the correlation function $\Pi (p)$
expressed using $c-$quark propagators and calculated in the operator product
expansion ($\mathrm{OPE}$) with some accuracy. In the case under discussion,
$\Pi ^{\mathrm{OPE}}(p)$ and corresponding amplitude $\Pi ^{\mathrm{OPE}%
}(p^{2})$ were computed in Ref.\ \cite{Agaev:2023wua}. There, we also found
the parameters $m_{0}$ and $f_{0}$ of the ground-state particle $X_{\mathrm{%
4c}}$, which appear in the present analysis as input quantities.

After the Borel transformation and continuum subtraction the SR equality
takes the form
\begin{equation}
f^{2}m^{2}e^{-m^{2}/M^{2}}=\Pi
(M^{2},s_{0})-f_{0}^{2}m_{0}^{2}e^{-m_{0}^{2}/M^{2}},  \label{eq:SR1}
\end{equation}%
which in conjunction with the derivation of Eq.\ (\ref{eq:SR1}) over $%
d/d(-1/M^{2})$, can be utilized to find sum rules for $m$ and $f$. Here, $%
\Pi (M^{2},s_{0})$ is the amplitude $\Pi ^{\mathrm{OPE}}(p^{2})$ after the
Borel transformation and subtraction operations, and $M^{2}$ and $s_{0}$ are
corresponding parameters.

The $\Pi (M^{2},s_{0})$ is given by the formula
\begin{equation}
\Pi (M^{2},s_{0})=\int_{16m_{c}^{2}}^{s_{0}}ds\rho ^{\mathrm{OPE}%
}(s)e^{-s/M^{2}}.  \label{eq:InvAmp}
\end{equation}%
where $\rho ^{\mathrm{OPE}}(s)$ is a two-point spectral density. It consists
of the perturbative contribution $\rho ^{\mathrm{pert.}}(s)$ and the
dimension-$4$ nonperturbative term $\sim \langle \alpha _{s}G^{2}/\pi
\rangle $: The explicit expression of $\rho ^{\mathrm{pert.}}(s)$ can be
found in Ref.\ \cite{Agaev:2023wua}.

To carry out numerical computations, one needs the gluon vacuum condensate $%
\langle \alpha _{s}G^{2}/\pi \rangle =(0.012\pm 0.004)~\mathrm{GeV}^{4}$ and
$c$-quark mass $m_{c}=(1.27\pm 0.02)~\mathrm{GeV}$. A crucial problem to be
clarified is a choice of the parameters $M^{2}$ and $s_{0}$. The regions in
which they can be changed should meet known restrictions of SR computations.
Stated differently, $M^{2}$ and $s_{0}$ have to be fixed in such a way that
to ensure dominance of the pole contribution ($\mathrm{PC}$) and
perturbative term over a nonperturbative one. The convergence of $\mathrm{OPE%
}$ and a stability of extracted observables against variations of the Borel
parameter $M^{2}$ are also among important constraints. Because, $\Pi
(M^{2},s_{0})$ does not contain quark and mixed condensates the dominance of
$\mathrm{PC}$ and stability of extracted quantities play key role in
choosing parameters $M^{2}$ and $s_{0}$.

\begin{widetext}

\begin{figure}[h!]
\begin{center}
\includegraphics[totalheight=6cm,width=8cm]{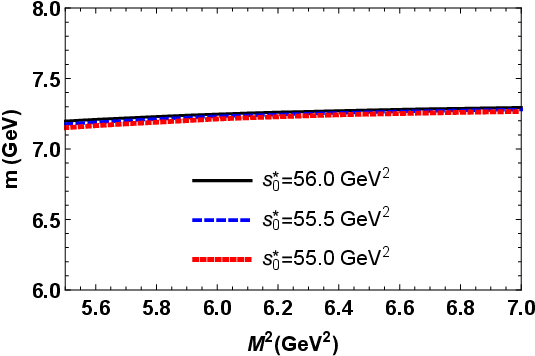}\,\, %
\includegraphics[totalheight=6cm,width=8cm]{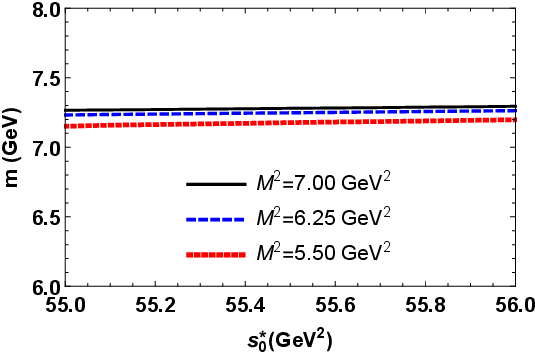}
\end{center}
\caption{ Mass of the tetraquark $X_{\mathrm{4c}^{\ast}}$ as a function of the
Borel $M^2$ (left), and
the continuum threshold $s_0^{\ast}$ parameters (right).}
\label{fig:Mass}
\end{figure}

\end{widetext}

In the first phase of computations, we fix the regions for $M^{2}$ and $%
s_{0} $ in order to activate in Eq.\ (\ref{eq:CorrF2}) only the ground-state
term. This task was fulfilled in Ref.\ \cite{Agaev:2023wua}, where $M^{2}$
and $s_{0}$ were varied inside the regions%
\begin{equation}
M^{2}\in \lbrack 5.5,7]~\mathrm{GeV}^{2},\ s_{0}\in \lbrack 49,50]~\mathrm{%
GeV}^{2}.  \label{eq:Wind1}
\end{equation}%
As a result, we evaluated the mass $m_{0}$ and coupling $f_{0}$ of the
ground-state tetraquark $X_{\mathrm{4c}}$
\begin{eqnarray}
m_{0} &=&(6570\pm 55)~\mathrm{MeV},  \notag \\
f_{0} &=&(5.61\pm 0.39)\times 10^{-2}~\mathrm{GeV}^{4}.  \label{eq:GState}
\end{eqnarray}%
At the second stage of studies, we use $m_{0}$ and $f_{0}$ in Eq.\ (\ref%
{eq:SR1}) as input parameters and calculate the mass $m$ and coupling $f$ of
the excited state
\begin{eqnarray}
m &=&(7235\pm 75)~\mathrm{MeV},  \notag \\
f &=&(8.0\pm 0.9)\times 10^{-2}~\mathrm{GeV}^{4}.  \label{eq:ExState}
\end{eqnarray}%
To compute Eq.\ (\ref{eq:ExState}), we use the working regions
\begin{equation}
M^{2}\in \lbrack 5.5,7]~\mathrm{GeV}^{2},\ s_{0}^{\ast }\in \lbrack 55,56]~%
\mathrm{GeV}^{2},  \label{eq:Region1}
\end{equation}%
which obey all constraints imposed on $\Pi (M^{2},s_{0})$ by the SR
analysis. In fact, the pole contribution changes inside limits $0.93\geq
\mathrm{PC}\geq 0.71$, at the minimum of $M^{2}=5.5~\mathrm{GeV}^{2}$ the
nonperturbative term is negative and constitutes only $1.4\%$ part of the
correlation function. The extracted quantities $m$ and coupling $f$ \ bear
residual dependence on the parameters $M^{2}$ and $s_{0}^{\ast }$ which is a
main source of theoretical uncertainties. These effects are equal to $\pm
1\% $ in the case of $m$, and to $\pm 11\%$ for $f$ staying within limits
acceptable for the SR computations. The behavior of the mass $m$ under
variations of $M^{2}$ and $s_{0}^{\ast }$ is shown in Fig.\ \ref{fig:Mass}.

Because we consider two terms in Eq.\ (\ref{eq:CorrF3}), and find parameters
of the ground-level and radially excited tetraquarks, there is a necessity
to check a self-consistency of performed studies. Indeed, the parameters $%
s_{0}$ and $s_{0}^{\ast }$ separate contributions of interest from ones
which are modeled using the assumption about quark-hadron duality.
Therefore, in these studies the inequalities $m_{0}^{2}<s_{0}$ and $%
s_{0}<m^{2}<s_{0}^{\ast }$ should be satisfied: With results of the
numerical analysis at hand, it is not difficult to verify these relations.

The prediction for the mass $m=7235~\mathrm{MeV}$ of the $2S$ excited
tetraquark $X_{\mathrm{4c}}^{\ast }$ within uncertainties of calculations
and errors of experiments is consistent with values $m^{\mathrm{ATL}%
}=7220\pm 30_{-30}^{+20}~\mathrm{MeV}$ and $m^{\mathrm{CMS}%
}=7287_{-18}^{+20}\pm 5~\mathrm{MeV}$, respectively. In our article \cite%
{Agaev:2023wua} we supposed that the resonance $X(7300)$ is $2S$ excited
state of $X(6600)$. This assumption based on the fact that the ATLAS
Collaboration detected the resonances $X(6600)$ and $X(7300)$ in $J/\psi
J/\psi $ and $J/\psi \psi ^{\prime }$ mass distributions, respectively.
Because \ the mass difference between mesons $\psi ^{\prime }$ and $J/\psi $
is around $590~\mathrm{MeV}$, and a comparable mass splitting $(600-735)~%
\mathrm{MeV}$ exists in the $X(7300)-X(6600)$ system, it is natural to
assume that $X(7300)$ is excitation of $X(6600)$. Our results for the masses
of $X_{\mathrm{4c}}$ and $X_{\mathrm{4c}}^{\ast }$ differ by amount $665~%
\mathrm{MeV}$ and seem support this scenario.


\subsection{The full width of $X_{\mathrm{4c}}^{\ast }$}


The mass $m$ of the excited tetraquark $X_{\mathrm{4c}}^{\ast }$ allow us to
determine its decay channels, and evaluate full width of this state. It is
clear, that decays to $J/\psi J/\psi $, $J/\psi \psi ^{\prime }$, $\eta
_{c}\eta _{c}$, $\eta _{c}\eta _{c}(2S)$, $\eta _{c}\chi _{c1}$, $\chi
_{c0}\chi _{c0}$, and $\chi _{c1}\chi _{c1}$ mesons are among such allowed
channels. It is worth noting that decay $X_{\mathrm{4c}}^{\ast }\rightarrow
\eta _{c}\chi _{c1}$ is the $P$-wave process, whereas the remaining ones are
$S$-wave decays.

We are going to explain in a detailed form only processes $X_{\mathrm{4c}%
}^{\ast }\rightarrow J/\psi J/\psi $ and $\ X_{\mathrm{4c}}^{\ast
}\rightarrow J/\psi \psi ^{\prime }$, and provide final results for other
channels. The partial widths of these decays are governed by the strong
couplings $g_{i}^{\ast }$ at the vertices $X_{\mathrm{4c}}^{\ast }J/\psi
J/\psi $, and $X_{\mathrm{4c}}^{\ast }J/\psi \psi ^{\prime }$. These
couplings can be evaluated using the following three-point correlation
function
\begin{eqnarray}
&&\Pi _{\mu \nu }(p,p^{\prime })=i^{2}\int d^{4}xd^{4}ye^{ip^{\prime
}y}e^{-ipx}\langle 0|\mathcal{T}\{J_{\mu }^{\psi }(y)  \notag \\
&&\times J_{\nu }^{\psi }(0)J^{\dagger }(x)\}|0\rangle ,  \label{eq:CorrF4}
\end{eqnarray}%
where $J_{\mu }^{\psi }(x)\ $ is the interpolating current for the mesons $%
J/\psi $ and $\psi ^{\prime }$
\begin{equation}
J_{\mu }^{\psi }(x)=\overline{c}_{i}(x)\gamma _{\mu }c_{i}(x),
\label{eq:Current2}
\end{equation}%
with $i=1,2,3$ being the color indices.

We apply usual recipes of the sum rule method and express the correlation
function $\Pi _{\mu \nu }(p,p^{\prime })$ in terms of physical parameters of
particles. Because the tetraquark $X_{\mathrm{4c}}^{\ast }$ decays both to $%
J/\psi J/\psi $ and $J/\psi \psi ^{\prime }$ pairs, we isolate in $\Pi _{\mu
\nu }(p,p^{\prime })$ contributions of the mesons $J/\psi $ and $\psi
^{\prime }$ from ones of higher resonances and continuum states. But the
current $J(x)$ also couples to the ground-state tetraquark $X_{\mathrm{4c}}$%
. Therefore, for the physical side of the sum rule $\Pi _{\mu \nu }^{\mathrm{%
Phys}}(p,p^{\prime })$, we get%
\begin{eqnarray}
&&\Pi _{\mu \nu }^{\mathrm{Phys}}(p,p^{\prime })=\sum_{I=1,2}\frac{\langle
0|J_{\mu }^{\psi }|J/\psi (p^{\prime })\rangle }{p^{\prime 2}-m_{J}^{2}}%
\frac{\langle 0|J_{\nu }^{\psi }|J/\psi (q)\rangle }{q^{2}-m_{J}^{2}}  \notag
\\
&&\times \langle J/\psi (p^{\prime })J/\psi (q)|X_{\mathrm{4c}%
}^{I}(p)\rangle \frac{\langle X_{\mathrm{4c}}^{I}(p)|J^{\dagger }|0\rangle }{%
p^{2}-m_{I}^{2}}  \notag \\
&&+\sum_{I=1,2}\frac{\langle 0|J_{\mu }^{\psi }|\psi (p^{\prime })\rangle }{%
p^{\prime 2}-m_{\psi }^{2}}\frac{\langle 0|J_{\nu }^{\psi }|J/\psi
(q)\rangle }{q^{2}-m_{J}^{2}}  \notag \\
&&\times \langle \psi (p^{\prime })J/\psi (q)|X_{\mathrm{4c}}^{I}(p)\rangle
\frac{\langle X_{\mathrm{4c}}^{I}(p)|J^{\dagger }|0\rangle }{p^{2}-m_{I}^{2}}%
\cdots ,  \label{eq:CorrF5}
\end{eqnarray}%
where $m_{J}=(3096.900\pm 0.006)~\mathrm{MeV}$ and $m_{\psi }=(3686.10\pm
0.06)~\mathrm{MeV}$ are the masses of $J/\psi $ and $\psi ^{\prime }$mesons
\cite{PDG:2022}. To write down $\Pi _{\mu \nu }^{\mathrm{Phys}}(p,p^{\prime
})$ in a compact form, we use in Eq.\ (\ref{eq:CorrF5}) notations $X_{%
\mathrm{4c}}^{1}=X_{\mathrm{4c}}$, $X_{\mathrm{4c}}^{2}=X_{\mathrm{4c}%
}^{\ast }$ and $m_{1}^{2}=m_{0}^{2}$, $m_{2}^{2}=m^{2}$.

The function $\Pi _{\mu \nu }^{\mathrm{Phys}}(p,p^{\prime })$ can be
expressed in terms of mesons and tetraquarks masses and decay constants
(couplings). To this end, one should use the matrix elements of the
tetraquarks Eq.\ (\ref{eq:MatEl1}), as well as the matrix elements
\begin{eqnarray}
\langle 0|J_{\mu }^{\psi }|J/\psi (p)\rangle &=&f_{J}m_{J}\varepsilon _{\mu
}(p),  \notag \\
\langle 0|J_{\mu }^{\psi }|\psi ^{\prime }(p)\rangle &=&f_{\psi }m_{\psi }%
\widetilde{\varepsilon }_{\mu }(p),  \label{eq:MatEl2}
\end{eqnarray}%
and
\begin{eqnarray}
&&\langle J/\psi (p^{\prime })J/\psi (q)|X_{\mathrm{4c}}(p)\rangle
=g_{1}(q^{2})\left[ q\cdot p^{\prime }\varepsilon ^{\ast }(p^{\prime })\cdot
\varepsilon ^{\ast }(q)\right.  \notag \\
&&\left. -q\cdot \varepsilon ^{\ast }(p^{\prime })p^{\prime }\cdot
\varepsilon ^{\ast }(q)\right] ,  \notag \\
&&\langle \psi (p^{\prime })J/\psi (q)|X_{\mathrm{4c}}(p)\rangle
=g_{2}(q^{2})\left[ q\cdot p^{\prime }\widetilde{\varepsilon }^{\ast
}(p^{\prime })\cdot \varepsilon ^{\ast }(q)\right.  \notag \\
&&\left. -q\cdot \widetilde{\varepsilon }^{\ast }(p^{\prime })p^{\prime
}\cdot \varepsilon ^{\ast }(q)\right] .  \label{eq:MatEl3}
\end{eqnarray}%
Here, $f_{J}=(409\pm 15)~\mathrm{MeV}$, $f_{\psi }=(279\pm 8)~\mathrm{MeV}$
and $\varepsilon _{\mu }$, $\widetilde{\varepsilon }_{\mu }$ are the decay
constants and polarization vectors of the mesons $J/\psi $ and $\psi
^{\prime }\ $\cite{PDG:2022,Kiselev:2001xa}, respectively. In the vertices
with the excited tetraquark $X_{\mathrm{4c}}^{\ast }(p)$ one should write
form factors $g_{1}^{\ast }(q^{2})$ and $g_{2}^{\ast }(q^{2})$.

Having used these matrix elements and carried out simple calculations, we
find for $\Pi _{\mu \nu }^{\mathrm{Phys}}(p,p^{\prime })$
\begin{eqnarray}
&&\Pi _{\mu \nu }^{\mathrm{Phys}}(p,p^{\prime
})=g_{1}(q^{2})f_{0}m_{0}f_{J}^{2}m_{J}^{2}F_{\mu \nu }(m_{0},m_{J})  \notag
\\
&&+g_{1}^{\ast }(q^{2})fmf_{J}^{2}m_{J}^{2}F_{\mu \nu }(m,m_{J})  \notag \\
&&+g_{2}(q^{2})f_{0}m_{0}f_{J}m_{J}f_{\psi }m_{\psi }F_{\mu \nu
}(m_{0},m_{\psi })+  \notag \\
&&+g_{2}^{\ast }(q^{2})fmf_{J}m_{J}f_{\psi }m_{\psi }F_{\mu \nu }(m,m_{\psi
})+\cdots ,  \label{eq:CorrF6}
\end{eqnarray}%
where
\begin{equation}
F_{\mu \nu }(a,b)=\frac{\left[ \left( a^{2}-b^{2}-q^{2}\right) g_{\mu \nu
}-2q_{\mu }p_{\nu }^{\prime }\right] }{2\left( p^{2}-a^{2}\right) \left(
p^{\prime 2}-b^{2}\right) (q^{2}-m_{J}^{2})}.
\end{equation}%
As is seen, there are two structures in $\Pi _{\mu \nu }^{\mathrm{Phys}%
}(p,p^{\prime })$ which can be used for SR analysis. To derive the sum rules
for the form factors $g_{i}^{(\ast )}(q^{2})$, we work with the Lorentz
structure $g_{\mu \nu }$, and corresponding invariant amplitude $\Pi ^{%
\mathrm{Phys}}(p^{2},p^{\prime 2},q^{2})$.

After the double Borel transformation of the function $\Pi ^{\mathrm{Phys}%
}(p^{2},p^{\prime 2},q^{2})$ over the variables $-p^{2}$ and $-p^{\prime 2}$%
, we get
\begin{eqnarray}
&&\mathcal{B}\Pi ^{\mathrm{Phys}}(p^{2},p^{\prime
2},q^{2})=g_{1}(q^{2})f_{0}m_{0}f_{J}^{2}m_{J}^{2}F(m_{0},m_{J})  \notag \\
&&+g_{1}^{\ast }(q^{2})fmf_{J}^{2}m_{J}^{2}F(m,m_{J})  \notag \\
&&+g_{2}(q^{2})f_{0}m_{0}f_{J}m_{J}f_{\psi }m_{\psi }F(m_{0},m_{\psi })
\notag \\
&&+g_{2}^{\ast }(q^{2})fmf_{J}m_{J}f_{\psi }m_{\psi }F(m,m_{\psi })+\cdots ,
\label{eq:CorrF7}
\end{eqnarray}%
with $F(a,b)$ being equal to
\begin{equation}
F(a,b)=\frac{\left( a^{2}-b^{2}-q^{2}\right) }{2(q^{2}-m_{J}^{2})}%
e^{-a^{2}/M_{1}^{2}}e^{-b^{2}/M_{2}^{2}}.
\end{equation}

The second component of the sum rules is the same correlation function $\Pi
_{\mu \nu }^{\mathrm{OPE}}(p,p^{\prime })$, but calculated using the $c$%
-quark propagators. The function $\Pi _{\mu \nu }^{\mathrm{OPE}}(p,p^{\prime
})$ and invariant amplitude $\Pi ^{\mathrm{OPE}}(p^{2},p^{\prime 2},q^{2})$
were computed in Ref.\ \cite{Agaev:2023wua}. Having equated $\mathcal{B}\Pi
^{\mathrm{Phys}}(p^{2},p^{\prime 2},q^{2})$ and the doubly Borel
transformation of the amplitude $\Pi ^{\mathrm{OPE}}(p^{2},p^{\prime
2},q^{2})$, and performed the continuum subtractions, we find the sum rule
equality, right-hand side of which is determined by the function
\begin{eqnarray}
&&\Pi (\mathbf{M}^{2},\mathbf{s}_{0},q^{2})=\int_{16m_{c}^{2}}^{s_{0}}ds%
\int_{4m_{c}^{2}}^{s_{0}^{\prime }}ds^{\prime }\rho (s,s^{\prime },q^{2})
\notag \\
&&\times e^{-s/M_{1}^{2}}e^{-s^{\prime }/M_{2}^{2}}.  \label{eq:SCoupl}
\end{eqnarray}%
where $\mathbf{M}^{2}=(M_{1}^{2},M_{2}^{2})$ and $\mathbf{s}%
_{0}=(s_{0},s_{0}^{\prime })$ are the Borel and continuum threshold
parameters, respectively. A spectral density $\rho (s,s^{\prime },q^{2})$ is
found as an imaginary part of $\Pi ^{\mathrm{OPE}}(p^{2},p^{\prime 2},q^{2})$%
. Let us note that parameters $(M_{1}^{2},\mathbf{s}_{0})$ and $%
(M_{2}^{2},s_{0}^{\prime })$ correspond to $X_{\mathrm{4c}}-X_{\mathrm{4c}%
}^{\ast }$ and $J/\psi -\psi ^{\prime }$ channels, respectively.

The equality Eq.\ (\ref{eq:CorrF7}) obtained by this way contains four
unknown form factors $g_{1(2)}^{(\ast )}(q^{2})$. One of possible methods to
extract them from this equality is to calculate its derivatives over $%
-1/M_{1}^{2}$ and $-1/M_{2}^{2}$. But then final expressions for $%
g_{1(2)}^{(\ast )}(q^{2})$ become rather complicated, which may reduce an
accuracy of numerical analyses. Here, we pursue the alternative policy: By
choosing appropriate subtraction parameters in $X_{\mathrm{4c}}-X_{\mathrm{4c%
}}^{\ast }$ and $J/\psi -\psi ^{\prime }$ channels, we include in analysis
terms from Eq.\ (\ref{eq:CorrF7}) one by one. These operations change number
of components in $\mathcal{B}\Pi ^{\mathrm{Phys}}$ and integration limits in
$\Pi (\mathbf{M}^{2},\mathbf{s}_{0},q^{2})$. At each new stage, we take into
account results obtained in previous steps, and solve subsequent equations
with only one unknown form factor.

First of all, let us note that the form factor $g_{1}(q^{2})$ was evaluated
in Ref.\ \cite{Agaev:2023wua}. It corresponds to the vertex $X_{\mathrm{4c}%
}J/\psi J/\psi $ and is necessary to compute the partial width of the decay $%
X_{\mathrm{4c}}\rightarrow J/\psi J/\psi $. To calculate $g_{1}(q^{2})$, we
fixed parameters $(M_{1}^{2},s_{0})$ as in Eq.\ (\ref{eq:Wind1}), whereas
for $(M_{2}^{2},s_{0}^{\prime })$ used
\begin{equation}
M_{2}^{2}\in \lbrack 4,5]~\mathrm{GeV}^{2}\text{, }s_{0}^{\prime }\in
\lbrack 12,13]~\mathrm{GeV}^{2},  \label{eq:Wind1A}
\end{equation}%
where $s_{0}^{\prime }$ is limited by the mass $m_{\psi }^{2}$ of the next
state in the $J/\psi -\psi ^{\prime }$ channel, i.e., $s_{0}^{\prime
}<m_{\psi }^{2}$. Afterwards, we choose $(M_{1}^{2},s_{0})$ in accordance
with Eq.\ (\ref{eq:Region1}), but do not modify $(M_{2}^{2},s_{0}^{\prime })$%
. By this way, we include into consideration $g_{1}^{\ast }(q^{2})$ and
obtain the equation containing $g_{1}(q^{2})$ and $g_{1}^{\ast }(q^{2})$ .
This means that remaining terms in Eq.\ (\ref{eq:CorrF7}) are included in
"higher resonances and continuum states" and their effects are implicitly
taken into account in $\Pi (\mathbf{M}^{2},\mathbf{s}_{0},q^{2})$ through
the quark-hadron duality. Then, using results for $g_{1}(q^{2})$, we
calculate the form factor $g_{1}^{\ast }(q^{2})$ that determines the width
of the process $X_{\mathrm{4c}}^{\ast }\rightarrow J/\psi J/\psi $.

At the new stage of studies, we consider the equation for the form factors $%
g_{1}(q^{2})$ and $g_{2}(q^{2})$. The latter corresponds to the vertex $X_{%
\mathrm{4c}}J/\psi \psi ^{\prime }$, and formally describes the channel $X_{%
\mathrm{4c}}\rightarrow J/\psi \psi ^{\prime }$. This decay mode of $X_{%
\mathrm{4c}}$ is kinematically forbidden, because the threshold $6737~%
\mathrm{MeV}$ for production of the $J/\psi \psi ^{\prime }$ pair exceeds
the mass of the tetraquark $X_{\mathrm{4c}}$. But $g_{2}(q^{2})$ is required
to determine the form factor $g_{2}^{\ast }(q^{2})$ of interest. To extract $%
g_{2}(q^{2})$, we fix $(M_{1}^{2},s_{0})$ by means of Eq.\ (\ref{eq:Wind1}),
but choose $(M_{2}^{2},s_{0}^{\ast \prime })$ in the form
\begin{equation}
M_{2}^{2}\in \lbrack 4,5]~\mathrm{GeV}^{2},\ s_{0}^{\ast \prime }\in \lbrack
15,16]~\mathrm{GeV}^{2},  \label{eq:Wind2}
\end{equation}%
where $s_{0}^{\ast \prime }<m_{\psi (3S)}^{2}$. Finally, using for the $X_{%
\mathrm{4c}}-X_{\mathrm{4c}}^{\ast }$ and $J/\psi -\psi ^{\prime }$ channels
Eqs.\ (\ref{eq:Region1}) and (\ref{eq:Wind2}), we calculate the last form
factor $g_{2}^{\ast }(q^{2})$.

The SR method allows one to calculate the form factors in the deep-Euclidean
region $q^{2}<0$. All functions $g_{i}^{(\ast )}(q^{2})$ in the present work
are calculated in the region $q^{2}=-(1-10)~\mathrm{GeV}^{2}$. But partial
widths of the decays under consideration are determined by values of these
form factors at the mass shell $q^{2}=m_{J}^{2}$. To solve this problem, we
introduce a new variable $Q^{2}=-q^{2}$ and denote the obtained functions by
$g_{i}^{(\ast )}(Q^{2})$. Afterwards, we use a fit functions $\mathcal{G}%
_{i}^{(\ast )}(Q^{2})$ that at momenta $Q^{2}>0$ are equal to the SR's
results, but can be extrapolated to the domain of $Q^{2}<0$. In present
article, we use functions $\mathcal{G}_{i}(Q^{2})$
\begin{equation}
\mathcal{G}_{i}(Q^{2})=\mathcal{G}_{i}^{0}\mathrm{\exp }\left[ c_{i}^{1}%
\frac{Q^{2}}{m^{2}}+c_{i}^{2}\left( \frac{Q^{2}}{m^{2}}\right) ^{2}\right] ,
\label{eq:FitF}
\end{equation}%
with parameters $\mathcal{G}_{i}^{0}$, $c_{i}^{1}$ and $c_{i}^{2}$. It is
worth noting that in the case of $g_{1}^{\ast }(q^{2})$ and $g_{2}^{\ast
}(q^{2})$ the parameter $m$ in Eq.\ (\ref{eq:FitF}) is the mass of the
tetraquark $X_{\mathrm{4c}}^{\ast }$, whereas for the intermediate functions
$g_{1}(q^{2})$ and $g_{2}(q^{2})$, we use the mass $m_{0}$ of $X_{\mathrm{4c}%
}$.

\begin{figure}[h]
\includegraphics[width=8.5cm]{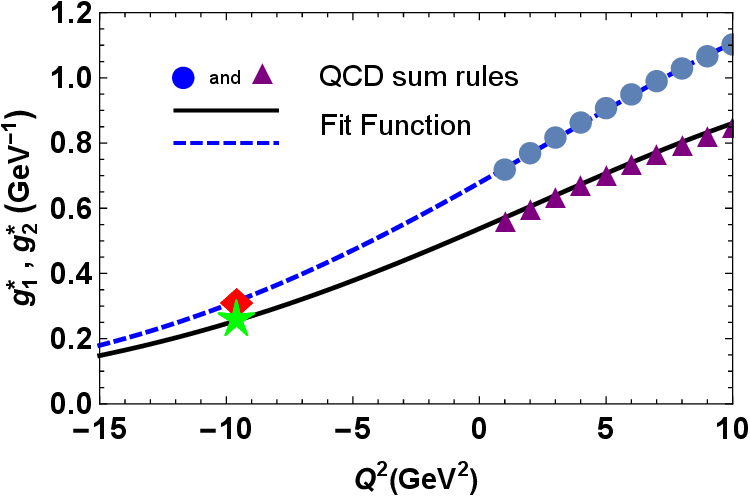}
\caption{The QCD results and fit functions for the form factors $g_{1}^{\ast
}(Q^{2})$ (dashed curve) and $g_{2}^{\ast }(Q^{2})$ (solid curve). The red
diamond and green star denote the point $Q^{2}=-m_{J}^{2}$, where the strong
couplings $g_{1}^{\ast }$ and $g_{2}^{\ast }$ are evaluated. }
\label{fig:Fit}
\end{figure}

Results obtained for $g_{1}^{\ast }(q^{2})$ and $g_{2}^{\ast }(q^{2})$ are
plotted in Fig.\ \ref{fig:Fit}. Computations demonstrate that $\mathcal{G}%
_{1}^{0\ast }=0.68~\mathrm{GeV}^{-1}$, $c_{1}^{1\ast }=3.93$, and $%
c_{1}^{2\ast }=-4.33$ lead to nice agreement with the sum rule's data for $%
g_{1}^{\ast }(Q^{2})$. At the mass shell $q^{2}=m_{J}^{2}$ the function $%
\mathcal{G}_{1}^{\ast }(Q^{2})$ is equal to
\begin{equation}
g_{1}^{\ast }\equiv \mathcal{G}_{1}^{\ast }(-m_{J}^{2})=(3.1\pm 0.5)\times
10^{-1}\ \mathrm{GeV}^{-1}.  \label{eq:Coupl1}
\end{equation}%
The width of the decay $X_{\mathrm{4c}}^{\ast }\rightarrow J/\psi J/\psi $
can be obtained by employing the expression
\begin{equation}
\Gamma \left[ X_{\mathrm{4c}}^{\ast }\rightarrow J/\psi J/\psi \right]
=g_{1}^{\ast 2}\frac{\lambda _{1}}{8\pi }\left( \frac{m_{J}^{4}}{m^{2}}+%
\frac{2\lambda _{1}^{2}}{3}\right) ,  \label{eq:PartDW}
\end{equation}%
where $\lambda _{1}=\lambda (m,m_{J},m_{J})$ and
\begin{eqnarray}
&&\lambda (m_{1},m_{2},m_{3})=\frac{\left[ m_{1}^{4}+m_{2}^{4}+m_{3}^{4}%
\right. }{2m_{1}}  \notag \\
&&\left. -2(m_{1}^{2}m_{2}^{2}+m_{1}^{2}m_{3}^{2}+m_{2}^{2}m_{3}^{2})\right]
^{1/2}.
\end{eqnarray}%
Then it is not difficult to find that
\begin{equation}
\Gamma \left[ X_{\mathrm{4c}}^{\ast }\rightarrow J/\psi J/\psi \right]
=(30.1\pm 8.3)~\mathrm{MeV}.  \label{eq:DW1}
\end{equation}

In the case of $g_{2}^{\ast }(Q^{2})$, similar investigations give for the
parameters of the function $\mathcal{G}_{2}^{\ast }(Q^{2})$ following
results: $\mathcal{G}_{2}^{0\ast }=0.54~\mathrm{GeV}^{-1}$, $c_{2}^{1\ast
}=3.28$, and $c_{2}^{2\ast }=-4.26$. The strong coupling $g_{2}^{\ast }$
equals to
\begin{equation}
g_{2}^{\ast }\equiv \mathcal{G}_{2}^{\ast }(-m_{J}^{2})=(2.5\pm 0.5)\times
10^{-1}\ \mathrm{GeV}^{-1}.
\end{equation}%
Partial width of the process $X_{\mathrm{4c}}^{\ast }\rightarrow J/\psi \psi
^{\prime }$ is given by the formula
\begin{equation}
\Gamma \left[ X_{\mathrm{4c}}^{\ast }\rightarrow J/\psi \psi ^{\prime }%
\right] =g_{2}^{\ast 2}\frac{\lambda _{2}}{8\pi }\left( \frac{m_{\psi
}^{2}m_{J}^{2}}{m^{2}}+\frac{2\lambda _{2}^{2}}{3}\right) ,
\end{equation}%
where $\lambda _{2}=\lambda (m,m_{\psi },m_{J})$. This leads to the
prediction
\begin{equation}
\Gamma \left[ X_{\mathrm{4c}}^{\ast }\rightarrow J/\psi \psi ^{\prime }%
\right] =(11.5\pm 3.3)~\mathrm{MeV}.  \label{eq:DW1A}
\end{equation}%
The results obtained for these two decay channels are collected in Table\ %
\ref{tab:Channels}.

The decays $X_{\mathrm{4c}}^{\ast }\rightarrow \eta _{c}\eta _{c}$ and $X_{%
\mathrm{4c}}^{\ast }\rightarrow \eta _{c}\eta _{c}(2S)$ can be explored in
the context of this scheme as well. In this case, the double Borel
transformation of the amplitude $\Pi _{\eta _{c}}^{\mathrm{Phys}%
}(p^{2},p^{\prime 2},q^{2})$ equals to
\begin{eqnarray}
&&\mathcal{B}\Pi _{\eta _{c}}^{\mathrm{Phys}}(p^{2},p^{\prime
2},q^{2})=g_{3}(q^{2})\frac{f_{0}m_{0}f_{\eta _{c}}^{2}m_{\eta _{c}}^{4}}{%
4m_{c}^{2}}R(m_{0},m_{\eta _{c}})  \notag \\
&&+g_{3}^{\ast }(q^{2})\frac{fmf_{\eta _{c}}^{2}m_{\eta _{c}}^{4}}{4m_{c}^{2}%
}R(m,m_{\eta _{c}})+g_{4}(q^{2})\frac{f_{0}m_{0}f_{\eta _{c}}m_{\eta
_{c}}^{2}}{4m_{c}^{2}}  \notag \\
&&\times f_{\eta _{c}(2S)}m_{\eta _{c}(2S)}^{2}R\left( m_{0},m_{\eta
_{c}(2S)}\right) +g_{4}^{\ast }(q^{2})\frac{fmf_{\eta _{c}}m_{\eta _{c}}^{2}%
}{4m_{c}^{2}}  \notag \\
&&\times f_{\eta _{c}(2S)}m_{\eta _{c}(2S)}^{2}R(m,m_{\eta _{c}(2S)})+\cdots
,  \label{eq:CorrF8}
\end{eqnarray}%
where $m_{\eta _{c}}=(2983.9\pm 0.4)~\mathrm{MeV}$, $f_{\eta _{c}}=(398.1\pm
1.0)~\mathrm{MeV}$ and $m_{\eta _{c}(2S)}=(3637.5\pm 1.1)~\mathrm{MeV}$, $%
f_{\eta _{c}(2S)}=331~\mathrm{MeV}$ are the spectroscopic parameters of the $%
\eta _{c}$ and $\eta _{c}(2S)$ mesons \cite{PDG:2022,Hatton:2020qhk}. The
function $R(a,b)$ is defined by the formula
\begin{equation}
R(a,b)=\frac{\left( a^{2}+b^{2}-q^{2}\right) }{2(q^{2}-m_{\eta _{c}}^{2})}%
e^{-a^{2}/M_{1}^{2}}e^{-b^{2}/M_{2}^{2}}.
\end{equation}

The invariant amplitude $\Pi _{\eta _{c}}^{\mathrm{OPE}}(p^{2},p^{\prime
2},q^{2})$ was calculated in our article \cite{Agaev:2023wua}. Here, one
should take into account that the regions $(M_{2}^{2},s_{0}^{\prime })$ and $%
(M_{2}^{2},s_{0}^{\ast \prime })$ for $\eta _{c}-\eta _{c}(2S)$ channel are
given by the expressions%
\begin{equation}
M_{2}^{2}\in \lbrack 3.5,4.5]~\mathrm{GeV}^{2},\ s_{0}^{\prime }\in \lbrack
11,12]~\mathrm{GeV}^{2},  \label{eq:Wind4}
\end{equation}%
and%
\begin{equation}
M_{2}^{2}\in \lbrack 3.5,4.5]~\mathrm{GeV}^{2},\ s_{0}^{\ast \prime }\in
\lbrack 13,14]~\mathrm{GeV}^{2},
\end{equation}%
respectively. In the case of $g_{3}^{\ast }(Q^{2})$, our studies lead for
the parameters of the function $\mathcal{G}_{3}^{\ast }(Q^{2})$ to
predictions: $\mathcal{G}_{3}^{0\ast }=0.39~\mathrm{GeV}^{-1}$, $%
c_{3}^{1\ast }=4.01$, and $c_{3}^{2\ast }=-4.99$. Then the coupling $%
g_{3}^{\ast }$ is equal to
\begin{equation}
g_{3}^{\ast }\equiv \mathcal{G}_{3}^{\ast }(-m_{\eta _{c}}^{2})=(1.7\pm
0.4)\times 10^{-1}\ \mathrm{GeV}^{-1}.
\end{equation}

The width of the decay $X_{\mathrm{4c}}^{\ast }\rightarrow \eta _{c}\eta
_{c} $ can be found by means of the formula%
\begin{equation}
\Gamma \left[ X_{\mathrm{4c}}^{\ast }\rightarrow \eta _{c}\eta _{c}\right]
=g_{3}^{\ast 2}\frac{m_{\eta _{c}}^{2}\lambda _{3}}{8\pi }\left( 1+\frac{%
\lambda _{3}^{2}}{m_{\eta _{c}}^{2}}\right) ,  \label{eq:PDw2}
\end{equation}%
where $\lambda _{3}=\lambda (m,m_{\eta _{c}},m_{\eta _{c}})$. Numerical
computations yield
\begin{equation}
\Gamma \left[ X_{\mathrm{4c}}^{\ast }\rightarrow \eta _{c}\eta _{c}\right]
=(30.6\pm 10.5)~\mathrm{MeV}.
\end{equation}%
For the second decay $X_{\mathrm{4c}}^{\ast }\rightarrow \eta _{c}\eta
_{c}(2S)$, we get
\begin{eqnarray}
&&g_{4}^{\ast }\equiv \mathcal{G}_{4}^{\ast }(-m_{\eta _{c}}^{2})=(1.4\pm
0.3)\times 10^{-1}\ \mathrm{GeV}^{-1},  \notag \\
&&\Gamma \left[ X_{\mathrm{4c}}^{\ast }\rightarrow \eta _{c}\eta _{c}(2S)%
\right] =(16.6\pm 5.5)~\mathrm{MeV},
\end{eqnarray}%
where $\mathcal{G}_{4}^{\ast }(Q^{2})$ is the function with parameters $%
\mathcal{G}_{4}^{0\ast }=0.32~\mathrm{GeV}^{-1}$, $c_{4}^{1\ast }=4.06$, and
$c_{4}^{2\ast }=-5.02$.

Treatment of the channels $X_{\mathrm{4c}}^{\ast }\rightarrow \eta _{c}\chi
_{c1}$, $\chi _{c0}\chi _{c0}$, and $\chi _{c1}\chi _{c1}$ is done by taking
into account vertices of the tetraquarks $X_{\mathrm{4c}}$ and $X_{\mathrm{4c%
}}^{\ast }$ with these meson pairs. Therefore, the physical side of the sum
rules consists of two terms. In the case of the $\eta _{c}\chi _{c1}$
mesons, both the ground-level tetraquark $X_{\mathrm{4c}}$ and its excited
state $X_{\mathrm{4c}}^{\ast }$ decays to this meson pair. Therefore, to
find the partial decay width of the process $X_{\mathrm{4c}}^{\ast
}\rightarrow \eta _{c}\chi _{c1}$, we use the form factor $g_{5}(q^{2})$
studied in Ref.\ \cite{Agaev:2023wua}, and extract $g_{5}^{\ast }(q^{2})$
necessary to compute the coupling $g_{5}^{\ast }$ at the mass shell $%
q^{2}=m_{\eta _{c}}^{2}$. The corresponding fit function $\mathcal{G}%
_{5}^{\ast }(Q^{2})$ has the parameters: $\mathcal{G}_{5}^{0\ast }=3.46$, $%
c_{5}^{1\ast }=3.59$, and $c_{5}^{2\ast }=-4.72$.

The remaining processes $X_{\mathrm{4c}}^{\ast }\rightarrow \chi _{c0}\chi
_{c0}$ and $\chi _{c1}\chi _{c1}$ are investigated by the same manner, the
difference being that decays of $X_{\mathrm{4c}}$ to mesons $\chi _{c0}\chi
_{c0}$, and $\chi _{c1}\chi _{c1}$ are not kinematically allowed channels,
but we compute relevant form factors to find strong couplings $g_{6}^{\ast }$
and $g_{7}^{\ast }$ of interests. The related correlation functions are
calculated in the present work for the first time and given by the
expressions (\ref{eq:ACF1}) and (\ref{eq:ACF2}). The final results of
analysis are collected in Table\ \ref{tab:Channels}. Let us note only that
in numerical computations, we employ the SR predictions for the decay
constants $f_{\chi _{c1}}=(344\pm 27)~\mathrm{MeV}$ and $f_{\chi _{c0}}=343~%
\mathrm{MeV}$ \cite{VeliVeliev:2012cc,Veliev:2010gb}.

Having used results for the partial widths of the excited $X_{\mathrm{4c}%
}^{\ast }$ tetraquark's decay channels, we estimate its full width
\begin{equation}
\Gamma =(144\pm 18)~\mathrm{MeV}.  \label{eq:ExcFW}
\end{equation}

\begin{widetext}

\begin{table}[tbp]
\begin{tabular}{|c|c|c|c|c|c|c|c|}
\hline\hline
$i$ & Channels &  $M_2^2 (\mathrm{GeV}^2)$ & $s_{0}^{(\ast)\prime} (\mathrm{GeV}^2)$ & $g_{i}^{\ast}\times10 ~(\mathrm{GeV}^{-1})$ & $\Gamma_{i}(\mathrm{MeV})$ & $G_{i}^{(\ast)}\times10~(\mathrm{GeV}^{-1})$ & $\widetilde {\Gamma}_{i}(\mathrm{MeV})$ \\ \hline
$1$ & $ J/\psi J/\psi$ &  $4-5$ & $12-13$ & $3.1 \pm 0.5$ & $30.1 \pm 8.3$ & $3.5 \pm 0.7$ & $36.1 \pm 10.5$ \\
$2$ & $J/\psi\psi^{\prime}$ &  $4-5$ & $15-16$ & $2.5 \pm 0.5$ & $11.5 \pm 3.3$ & $3.2 \pm 0.5$ & $16.2 \pm 5.1$ \\
$3$ & $\eta_{c}\eta_{c}$ &  $3.5-4.5$ & $11-12$ & $1.7 \pm 0.4$ & $30.6 \pm 10.5$ & $2.0 \pm 0.4$ & $42.3 \pm 12.2$ \\
$4$ & $\eta_{c}\eta_{c}(2S)$ & $3.5-4.5$ & $13-14$ & $1.4 \pm 0.3$ & $16.6 \pm 5.5$ & $1.3 \pm 0.3$ & $14.7 \pm 5.1$ \\
$5$ & $\eta_{c}\chi_{c1}$ &  $4-5$ & $13-14$ & $16.4 \pm 3.8$ & $11.6 \pm 4.1$ & $18.3 \pm 4.1$ & $12.8 \pm 4.2$ \\
$6$ & $\chi_{c0}\chi_{c0}$ &  $4-5$ & $14-14.9$ & $2.1 \pm 0.4$ & $28.8 \pm 7.9$ & $2.3 \pm 0.5$ & $29.9 \pm 9.4$ \\
$7$ & $ \chi_{c1}\chi_{c1}$ &   $4-5$ & $13-14$ & $3.5 \pm 0.5$ & $14.4 \pm 4.2$ & $4.1 \pm 0.8$ & $16.5 \pm 4.7$\\
 \hline\hline
\end{tabular}%
\caption{Decay channels of the tetraquark $X_{\mathrm{4c}}^{\ast}$ and hadronic molecule $\mathcal{M}$,
strong couplings $g_{i}^{\ast}$ and $G_{i}^{(\ast)}$, and partial widths $\Gamma_{i}$ and $\widetilde{\Gamma}_{i}$. The couplings $g_5^{\ast}$ and $G_5$ are dimensionless. For all decays of the tetraquark $X_{\mathrm{4c}}^{\ast}$ the parameters $M_{1}^{2}$ and $s_{0}^{\ast }$ vary in the regions $[5.5,7]~\mathrm{GeV}^{2}$ and $[55,56]~\mathrm{GeV}^{2}$, respectively. In the case of the molecule $\mathcal{M}$ the parameters are $M_{1}^{2}\in \lbrack 6,8]~\mathrm{GeV}^{2}$ and $s_{0}\in \lbrack 63,65]~\mathrm{GeV}^{2}$ for all considered processes. }
\label{tab:Channels}
\end{table}

\end{widetext}


\section{Hadronic molecule $\protect\chi _{c1}\protect\chi _{c1}$}

\label{sec:Molecule}

Here, we investigate the hadronic molecule $\mathcal{M}=\chi _{c1}\chi _{c1}$
and calculate the mass and current coupling of this structure, which will be
used to determine its kinematically allowed decay channels. Decays of the
molecule $\mathcal{M}$ and its full width are also studied in this section.


\subsection{Mass and current coupling}


The sum rules for the mass $\widetilde{m}$ and current coupling $\widetilde{f%
}$ of the molecule $\mathcal{M}$ can be extracted by exploring the
correlation function
\begin{equation}
\Pi (p)=i\int d^{4}xe^{ipx}\langle 0|\mathcal{T}\{\widetilde{J}(x)\widetilde{%
J}^{\dag }(0)\}|0\rangle .  \label{eq:CorrF9}
\end{equation}%
Here, $\widetilde{J}(x)$ is the interpolating current for $\mathcal{M}$%
\begin{equation}
\widetilde{J}(x)=\overline{c}_{a}(x)\gamma _{5}\gamma _{\mu }c_{a}(x)%
\overline{c}_{b}(x)\gamma _{5}\gamma ^{\mu }c_{b}(x),  \label{eq:Corr1}
\end{equation}%
with $a$, and $b$ being the color indices. We are going to calculate
spectroscopic parameters of the ground-level molecule $\mathcal{M}$,
therefore the physical side of the SRs is given by only one term
\begin{equation}
\Pi ^{\mathrm{Phys}}(p)=\frac{\widetilde{f}^{2}\widetilde{m}^{2}}{\widetilde{%
m}^{2}-p^{2}}+\cdots .  \label{eq:CorrF10}
\end{equation}%
It is calculated by taking into account the matrix element
\begin{equation}
\langle 0|\widetilde{J}|\mathcal{M}\rangle =\widetilde{f}\widetilde{m}.
\label{eq:MatEl4}
\end{equation}%
The invariant amplitude that is required for following analysis is $\Pi ^{%
\mathrm{Phys}}(p^{2})=\widetilde{f}^{2}\widetilde{m}^{2}/(\widetilde{m}%
^{2}-p^{2})$.

The correlation function $\Pi ^{\mathrm{OPE}}(p)$ in terms of the $c$-quark
propagators is determined by Eq.\ (\ref{eq:A1})%
\begin{eqnarray}
&&\Pi ^{\mathrm{OPE}}(p)=i\int d^{4}xe^{ipx}\left\{ \mathrm{Tr}\left[ \gamma
_{5}\gamma _{\mu }S_{c}^{ba^{\prime }}(x)\gamma _{\nu }\gamma
_{5}S_{c}^{a^{\prime }b}(-x)\right] \right.  \notag \\
&&\times \mathrm{Tr}\left[ \gamma _{5}\gamma ^{\mu }S_{c}^{ab^{\prime
}}(x)\gamma ^{\nu }\gamma _{5}S_{c}^{b^{\prime }a}(-x)\right] -\mathrm{Tr}%
\left[ \gamma _{5}\gamma _{\mu }S_{c}^{bb^{\prime }}(x)\gamma _{\nu }\right.
\notag \\
&&\left. \times \gamma _{5}S_{c}^{b^{\prime }a}(-x)\gamma _{5}\gamma ^{\mu
}S_{c}^{aa^{\prime }}(x)\gamma ^{\nu }\gamma _{5}S_{c}^{a^{\prime }b}(-x)
\right] -\mathrm{Tr}\left[ \gamma _{5}\gamma _{\mu }\right.  \notag \\
&&\left. \times S_{c}^{ba^{\prime }}(x)\gamma _{\nu }\gamma
_{5}S_{c}^{a^{\prime }a}(-x)\gamma _{5}\gamma ^{\mu }S_{c}^{ab^{\prime
}}(x)\gamma ^{\nu }\gamma _{5}S_{c}^{b^{\prime }b}(-x)\right]  \notag \\
&&+\mathrm{Tr}\left[ \gamma _{5}\gamma _{\mu }S_{c}^{bb^{\prime }}(x)\gamma
_{\nu }\gamma _{5}S_{c}^{b^{\prime }b}(-x)\right] \mathrm{Tr}\left[ \gamma
_{5}\gamma ^{\mu }S_{c}^{aa^{\prime }}(x)\gamma ^{\nu }\right.  \notag \\
&&\left. \left. \times \gamma _{5}S_{c}^{a^{\prime }a}(-x)\right] \right\} .
\label{eq:A1}
\end{eqnarray}%
It is convenient to denote the invariant amplitude of the QCD side by $\Pi ^{%
\mathrm{OPE}}(p^{2})$. Then, the sum rules for the mass and current coupling
take simple forms
\begin{equation}
\widetilde{m}^{2}=\frac{\Pi ^{\prime }(M^{2},s_{0})}{\Pi (M^{2},s_{0})}
\label{eq:Mass}
\end{equation}%
and \
\begin{equation}
\widetilde{f}^{2}=\frac{e^{\widetilde{m}^{2}/M^{2}}}{\widetilde{m}^{2}}\Pi
(M^{2},s_{0}),  \label{eq:Coupl}
\end{equation}%
where $\Pi ^{\prime }(M^{2},s_{0})=d\Pi (M^{2},s_{0})/d(-1/M^{2})$. Here, $%
\Pi (M^{2},s_{0})$ is the amplitude $\Pi ^{\mathrm{OPE}}(p^{2})$ obtained
after the Borel transformation and continuum subtraction operations.

\begin{figure}[h]
\includegraphics[width=8.5cm]{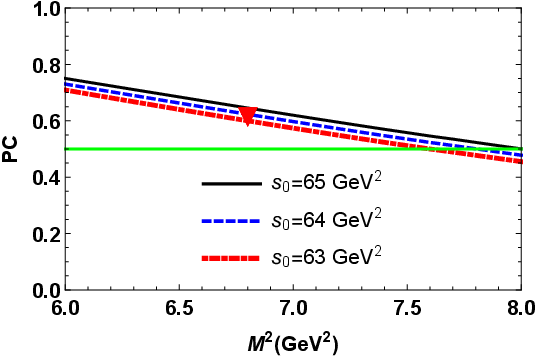}
\caption{Dependence of $\mathrm{PC}$ on the Borel parameter $M^{2}$. The
horizontal line shows the border $\mathrm{PC}=0.5$. The red triangle fix the
position, where the mass of the molecule $\protect\chi _{c1}\protect\chi %
_{c1}$ has been evaluated.}
\label{fig:PC}
\end{figure}

Computations lead to the following constraints on the parameters $M^{2}$ and
$s_{0}$%
\begin{equation}
M^{2}\in \lbrack 6,8]~\mathrm{GeV}^{2},\ s_{0}\in \lbrack 63,65]~\mathrm{GeV}%
^{2}.  \label{eq:Wind5}
\end{equation}%
It is not difficult to check that $\mathrm{PC}$ meets usual requirements of
SR computations. In Fig.\ \ref{fig:PC}, we plot dependence of the pole
contribution on the Borel parameter. It is seen, that expect for a small
region, $\mathrm{PC}$ is larger than $0.5.$ On average in $s_{0}$ the $%
\mathrm{PC}$ exceeds $0.5$ for all values of $M^{2}$.

The mass and current coupling of the molecule $\mathcal{M}$ are
\begin{eqnarray}
\widetilde{m} &=&(7180\pm 120)~\mathrm{MeV},  \notag \\
\widetilde{f} &=&(1.06\pm 0.13)\times 10^{-1}~\mathrm{GeV}^{4},
\label{eq:Results1}
\end{eqnarray}%
respectively. It is worth to note that $\widetilde{m}$ and $\widetilde{f}$
in Eq.\ (\ref{eq:Results1}) are mean values of the mass and current coupling
averaged over the working regions (\ref{eq:Wind5}). It overshoots the mass $%
7022~\mathrm{MeV}$ of two $\chi _{c1}$ mesons by $160~\mathrm{MeV}$ and is
unstable against decays to these particles.

\begin{widetext}

\begin{figure}[h!]
\begin{center}
\includegraphics[totalheight=6cm,width=8cm]{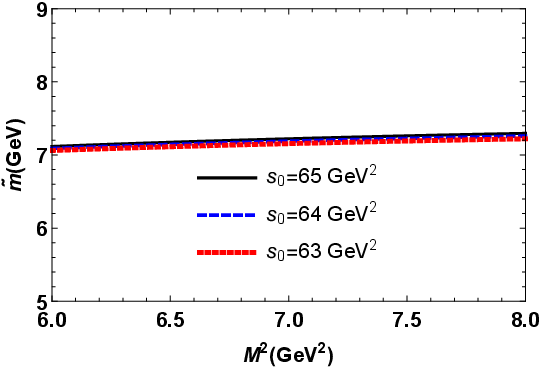}\,\, %
\includegraphics[totalheight=6cm,width=8cm]{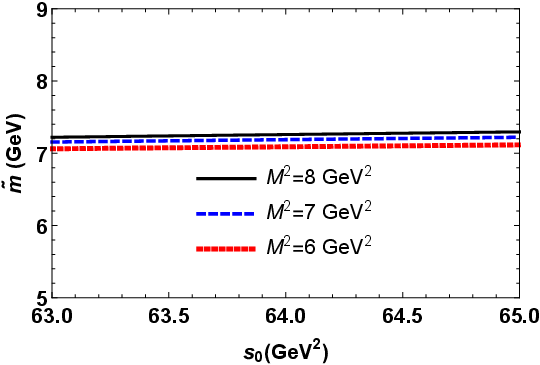}
\end{center}
\caption{ Mass $\widetilde{m}$ of the molecule $\chi_{c1}\chi_{c1}$.}
\label{fig:MassB}
\end{figure}

\end{widetext}

In Fig.\ \ref{fig:MassB}, we plot the mass $\widetilde{m}$ as a function of $%
M^{2}$ and $s_{0}$, in which its residual dependence on these parameters is
clear. It is also useful to estimate a gap between the ground-state $%
\mathcal{M}$ and excited molecules $\mathcal{M}^{\ast }$. The mass $%
\widetilde{m}^{\ast }$ of the state $\mathcal{M}^{\ast }$ should obey the
constraint $\widetilde{m}^{\ast }\geq \sqrt{s_{0}}$, i.e., $\widetilde{m}%
^{\ast }\geq 8~\mathrm{GeV}$, which implies an approximately $800~\mathrm{MeV%
}$ mass splitting between these molecules.

\subsection{Width of $\mathcal{M}$}


Decay channels of the hadronic molecule $\mathcal{M}$ do not differ from
that of the tetraquark $X_{\mathrm{4c}}^{\ast }$. A difference appears in
treatment of these processes. Indeed, the molecule $\mathcal{M}$ is
ground-state particle it its class, therefore physical side of relevant sum
rules in $\mathcal{M}$ channel contains terms connected only with its decays.

Because the resonances under investigation were detected in the di-$J/\psi $
and $J/\psi \psi ^{\prime }$ mass distributions, we concentrate on the
decays $\mathcal{M}\to J/\psi J/\psi $ and $\mathcal{M}\to J/\psi \psi
^{\prime }$. The correlation function required for this analysis is given by
the formula
\begin{eqnarray}
\widetilde{\Pi }_{\mu \nu }(p,p^{\prime }) &=&i^{2}\int
d^{4}xd^{4}ye^{ip^{\prime }y}e^{-ipx}\langle 0|\mathcal{T}\{J_{\mu }^{\psi
}(y)  \notag \\
&&\times J_{\nu }^{\psi }(0)\widetilde{J}^{\dagger }(x)\}|0\rangle .
\label{eq:CorrF11}
\end{eqnarray}%
As usual, we express $\widetilde{\Pi }_{\mu \nu }(p,p^{\prime })$ in terms
of the physical parameters of particles involved in the decay process. To
this end, we write it in the following form
\begin{eqnarray}
&&\widetilde{\Pi }_{\mu \nu }^{\mathrm{Phys}}(p,p^{\prime })=\frac{\langle
0|J_{\mu }^{\psi }|J/\psi (p^{\prime })\rangle }{p^{\prime 2}-m_{J}^{2}}%
\frac{\langle 0|J_{\nu }^{\psi }|J/\psi (q)\rangle }{q^{2}-m_{J}^{2}}  \notag
\\
&&\times \langle J/\psi (p^{\prime })J/\psi (q)|\mathcal{M}(p)\rangle \frac{%
\langle \mathcal{M}(p)|\widetilde{J}^{\dagger }|0\rangle }{p^{2}-\widetilde{m%
}^{2}}  \notag \\
&&+\frac{\langle 0|J_{\mu }^{\psi }|\psi (p^{\prime })\rangle }{p^{\prime
2}-m_{\psi }^{2}}\frac{\langle 0|J_{\nu }^{\psi }|J/\psi (q)\rangle }{%
q^{2}-m_{J}^{2}}  \notag \\
&&\times \langle \psi (p^{\prime })J/\psi (q)|\mathcal{M}(p)\rangle \frac{%
\langle \mathcal{M}(p)|\widetilde{J}^{\dagger }|0\rangle }{p^{2}-\widetilde{m%
}^{2}}+\cdots .  \label{eq:CorrF12}
\end{eqnarray}%
We have already defined the matrix elements of the hadronic molecule $%
\mathcal{M}$ and mesons $J/\psi $ and $\psi ^{\prime }$. The vertices $%
\mathcal{M}J/\psi J/\psi $ and $\mathcal{M}J/\psi \psi ^{\prime }$ after
some substitutions are given by Eq.\ (\ref{eq:MatEl3}). As in previous
section, we use the amplitude $\widetilde{\Pi }^{\mathrm{Phys}%
}(p^{2},p^{\prime 2},q^{2})$ which in $\widetilde{\Pi }_{\mu \nu }^{\mathrm{%
Phys}}(p,p^{\prime })$ corresponds to a term proportional to $g_{\mu \nu }$.
The double Borel transformation of the function $\widetilde{\Pi }^{\mathrm{%
Phys}}(p^{2},p^{\prime 2},q^{2})$ over the variables $-p^{2}$ and $%
-p^{\prime 2}$ is equal to
\begin{eqnarray}
&&\mathcal{B}\widetilde{\Pi }^{\mathrm{Phys}}(p^{2},p^{\prime
2},q^{2})=G_{1}(q^{2})\widetilde{f}\widetilde{m}f_{J}^{2}m_{J}^{2}F(%
\widetilde{m},m_{J})  \notag \\
&&+G_{1}^{\ast }(q^{2})\widetilde{f}\widetilde{m}f_{J}m_{J}f_{\psi }m_{\psi
}F(\widetilde{m},m_{\psi })+\cdots .  \label{eq:CorrF13}
\end{eqnarray}%
The correlation function $\widetilde{\Pi }_{\mu \nu }^{\mathrm{OPE}%
}(p,p^{\prime })$ is given by the formula
\begin{eqnarray}
&&\widetilde{\Pi }_{\mu \nu }^{\mathrm{OPE}}(p,p^{\prime })=2i^{2}\int
d^{4}xd^{4}ye^{-ipx}e^{ip^{\prime }y}\left\{ \mathrm{Tr}\left[ \gamma _{\nu
}S_{c}^{jb}(-x)\right. \right.  \notag \\
&&\left. \times \gamma ^{\alpha }\gamma _{5}S_{c}^{bj}(x)\right] \mathrm{Tr}%
\left[ \gamma _{\mu }S_{c}^{ia}(y-x)\gamma _{\alpha }\gamma
_{5}S_{c}^{ai}(x-y)\right]  \notag \\
&&-\mathrm{Tr}\left[ \gamma _{\mu }S_{c}^{ia}(y-x)\gamma _{\alpha }\gamma
_{5}S_{c}^{aj}(x)\gamma _{\nu }S_{c}^{jb}(-x)\gamma ^{\alpha }\right.  \notag
\\
&&\left. \left. \times \gamma _{5}S_{c}^{bi}(x-y)\right] \right\} .
\label{eq:A2}
\end{eqnarray}%
The QCD side of the sum rule and amplitude $\widetilde{\Pi }^{\mathrm{OPE}%
}(p^{2},p^{\prime 2},q^{2})$ are extracted from this expression. The
strategy pursued in our study of these processes repeats one used in Sec.\ %
\ref{sec:Excitation} while considering decays of the tetraquark $X_{\mathrm{%
4c}}^{\ast }$. We first determine the form factor $G_{1}(q^{2})$ utilizing
the "ground-state + continuum" scheme. The parameters $(M_{1}^{2},s_{0})$
are universal for all decays of $\mathcal{M}$ and are presented in Eq.\ (\ref%
{eq:Wind5}). The second pair of the parameters $(M_{2}^{2},s_{0}^{\prime })$
corresponding to $J/\psi J/\psi $ decay can be found in Eq.\ (\ref{eq:Wind1A}%
). Once determined $G_{1}(q^{2})$, in the second stage of computations we
choose $(M_{2}^{2},s_{0}^{\ast \prime })$ from Eq.\ (\ref{eq:Wind2}) and
employ information on $G_{1}(q^{2})$ to find the form factor $G_{1}^{\ast
}(q^{2})$, responsible for the process $\mathcal{M}\to J/\psi \psi ^{\prime
} $. The functions $\mathcal{G}_{8}(Q^{2})$ and $\mathcal{G}_{8}^{\ast
}(Q^{2}) $ are formed by the parameters%
\begin{eqnarray}
\mathcal{G}_{8}^{0} &=&0.76~\mathrm{GeV}^{-1}\ ,\ c_{8}^{1}=3.32,\
c_{8}^{2}=-4.19,  \notag \\
\mathcal{G}_{8}^{0\ast } &=&0.68~\mathrm{GeV}^{-1},c_{8}^{1\ast }=3.20,\
c_{8}^{2\ast }=-4.11.
\end{eqnarray}%
The strong couplings $G_{1}$ and $G_{1}^{\ast }$ are extracted from these
functions at the mass shells $Q^{2}=-m_{J}^{2}$.

This approach is also valid for the channels $\mathcal{M}\rightarrow \eta
_{c}\eta _{c}$ and $\mathcal{M}\rightarrow \eta _{c}\eta _{c}(2S)$. The
correlation function required for these decays is written down below
\begin{eqnarray}
&&\Pi ^{\mathrm{OPE}}(p,p^{\prime })=2\int d^{4}xd^{4}ye^{-ipx}e^{ip^{\prime
}y}\left\{ \mathrm{Tr}\left[ \gamma _{5}S_{c}^{ia}(y-x)\right. \right.
\notag \\
&&\left. \times \gamma _{\alpha }\gamma _{5}S_{c}^{ai}(x-y)\right] \mathrm{Tr%
}\left[ \gamma _{5}S_{c}^{jb}(-x)\gamma ^{\alpha }\gamma _{5}S_{c}^{bj}(x)%
\right]  \notag \\
&&-\mathrm{Tr}\left[ \gamma _{5}S_{c}^{ia}(y-x)\gamma _{\alpha }\gamma
_{5}S_{c}^{aj}(x)\gamma _{5}S_{c}^{jb}(-x)\gamma ^{\alpha }\right.  \notag \\
&&\left. \left. \times \gamma _{5}S_{c}^{bi}(x-y)\right] \right\} .  \label{eq:A3}
\end{eqnarray}%
The functions $\mathcal{G}_{9}(Q^{2})$ and $\mathcal{G}_{9}^{\ast }(Q^{2})$
needed to extrapolate the form factors $G_{2}(q^{2})$ and $G_{2}^{\ast
}(q^{2})$ are determined by the parameters: $\mathcal{G}_{9}^{0}=0.46~%
\mathrm{GeV}^{-1}\ ,\ c_{9}^{1}=3.93,\ c_{9}^{2}=-4.83$ and $\mathcal{G}%
_{9}^{0\ast }=0.30~\mathrm{GeV}^{-1},c_{9}^{1\ast }=3.90,\ c_{9}^{2\ast
}=-4.81$. These functions at the mass shells $Q^{2}=-m_{J}^{2}$ fix the
couplings $G_{2}$ and $G_{2}^{\ast }$, respectively.

The decays $\mathcal{M}\rightarrow \eta _{c}\chi _{c1}$, $\chi _{c0}\chi
_{c0}$, and $\chi _{c1}\chi _{c1}$ are investigated directly in the context
of the "ground-state + continuum" approach. Corresponding functions $\Pi
_{\mu }^{\mathrm{OPE}}(p,p^{\prime })$, $\Pi ^{\mathrm{OPE}}(p,p^{\prime })$
and $\widehat{\Pi }_{\mu \nu }^{\mathrm{OPE}}(p,p^{\prime })$ can found in
Appendix as Eqs.\ (\ref{eq:A4}) -(\ref{eq:A6}). Predictions obtained for the
partial widths of different modes of the hadronic molecule $\mathcal{M}$,
strong couplings and related parameters are presented in Table \ref%
{tab:Channels}. It should be noted that, to collect results obtained in this
work in the framework of a single Table, the couplings $G_{1}^{\ast }$, $%
G_{2}$ and $G_{2}^{\ast }$ are placed there under numbers $G_{2}^{\ast }$, $%
G_{3}$ and $G_{4}^{\ast }$, respectively.

For the full width of the hadronic molecule, we get
\begin{equation}
\widetilde{\Gamma }=(169\pm 21)~\mathrm{MeV},
\end{equation}%
which characterizes it as a wide structure.


\section{Summing up}

\label{sec:Disc}

In the present work, we have explored radially excited tetraquark $X_{%
\mathrm{4c}}^{\ast }$ and hadronic molecule $\mathcal{M}=\chi _{c1}\chi
_{c1} $. We have computed their masses and full widths using the QCD sum
rule method and interpolating currents $J(x)$ and $\widetilde{J}(x)$.
Obtained results have been confronted with available data of the ATLAS-CMS
Collaborations on the heaviest resonance $X(7300)$.

The LHCb fixed this state at $7.2~\mathrm{GeV}$, but did not provide other
information. The CMS measured parameters of this resonance and found that%
\begin{eqnarray}
m^{\mathrm{CMS}} &=&7287_{-18}^{+20}\pm 5~\mathrm{MeV},  \notag \\
\Gamma ^{\mathrm{CMS}} &=&95_{-40}^{+59}\pm 19~\mathrm{MeV}.  \label{eq:CMS}
\end{eqnarray}%
The ATLAS Collaboration observed $X(7300)$ in the $J/\psi \psi ^{\prime }$
mass distribution and also reported the mass and width of this state

\begin{eqnarray}
m^{\mathrm{ATL}} &=&7220\pm 30_{-30}^{+20}~\mathrm{MeV},  \notag \\
\Gamma ^{\mathrm{ATL}} &=&100_{-70-50}^{+130+60}~\mathrm{MeV}.
\label{eq:ATLAS}
\end{eqnarray}%
As is seen, experimental data suffer from big errors: only in the case of
Eq.\ (\ref{eq:CMS}) they are relatively small.

Comparing our findings $m=(7235\pm 75)~\mathrm{MeV}$ and $\widetilde{m}%
=(7180\pm 120)~\mathrm{MeV}$ with corresponding experimental data and taking
into account errors of calculations and measurements, we conclude that
masses of the excited tetraquark $X_{\mathrm{4c}}^{\ast }$ and hadronic
molecule $\mathcal{M}$ are compatible with $m^{\mathrm{CMS}}$ and $m^{%
\mathrm{ATL}}$. In other words, at this phase of analysis, it is difficult
to make assignment for the resonance $X(7300)$.

The full widths of the structures $X_{\mathrm{4c}}^{\ast }$ and $\mathcal{M}$
provide very important information for this purpose. It is interesting that $%
X(7300)$ is narrowest fully charmed state detected by the ATLAS and CMS
experiments provided one ignores errors of measurements. The four-quark
structures $X_{\mathrm{4c}}^{\ast }$ and $\mathcal{M}$ have the widths $%
\Gamma =(144\pm 18)~\mathrm{MeV}$ and $\widetilde{\Gamma }=(169\pm 21)~%
\mathrm{MeV}$, respectively. As is seen, within uncertainties of theoretical
analysis they agree with results of measurements. Because masses of these
structures are also consistent with existing data, both the excited
tetraquark $X_{\mathrm{4c}}^{\ast }$ and hadronic molecule $\mathcal{M}$ may
be considered as natural candidates to the observed state $X(7300)$.

For more strong conclusions about internal organization of $X(7300)$, it is
useful to examine an overlap of the currents $J(x)$ and $\widetilde{J}(x)$
with the physical state $X(7300)$ modeled as the structures $X_{\mathrm{4c}%
}^{\ast }$ or $\mathcal{M}$. This information is encoded in the matrix
element
\begin{equation}
\langle 0|J|X(7300)\rangle =\Lambda _{J}.  \label{eq:Overlap}
\end{equation}%
By employing Eqs.\ (\ref{eq:ExState}) and (\ref{eq:Results1}) obtained for
the diquark-antidiquark and molecule currents, we find that the matrix
element in Eq.\ (\ref{eq:Overlap}) equals to $\Lambda _{J}\approx 0.58~\mathrm{%
GeV}^{5}$ and $\Lambda _{\widetilde{J}}\approx 0.76~\mathrm{GeV}^{5}$,
respectively. This means, that the fully charmed resonance $X(7300)$ couples
with a larger strength to $\chi _{c1}\chi _{c1}$ molecule current \ than to
the current from Eq.\ (\ref{eq:Current1}). But from the ratio $\Lambda
_{J}/\Lambda _{\widetilde{J}}\approx 0.76$ it also becomes clear that $%
X(7300)$ can not be interpreted as a pure molecule state. Indeed, the
diquark-antidiquark current $J(x)$ through Fierz transformations can be
expressed as a weighted sum of  $\widetilde{J}(x)$ and other currents. Then
if $X(7300)$ had a pure molecule structure, the ratio $\Lambda _{J}/\Lambda
_{\widetilde{J}}$ would be equal to a weight of the molecule component in $%
J(x)$, and considerably smaller than $0.76$ \cite{Nielsen:2009uh}. Because
this is not a case, we can conclude that $X(7300)$\ contains a sizeable $X_{%
\mathrm{4c}}^{\ast }$ component. As a result, a preferable model for the
resonance $X(7300)$ is the admixture of the molecule $\mathcal{M}$ with
considerable piece of the tetraquark $X_{\mathrm{4c}}^{\ast }$.

Parameters of such mixing depend on precision of the mass and width of the
resonance $X(7300)$ measured by experimental groups. Accuracy of the
theoretical results are also important. In the sum rule method physical
observables are evaluated with some accuracy and contain uncertainties which
can be kept under control. The ambiguities in the masses and widths of the
structures $X_{\mathrm{4c}}^{\ast }$ and $\mathcal{M}$ are typical for such
kind of investigations, and can hardly be reduced. Therefore, for
quantitative analysis of the $\mathcal{M-}X_{\mathrm{4c}}^{\ast }$ mixing
phenomenon one needs more precise experimental data. This is true not only
for $X(7300)$, but also for other fully charmed $X$ resonances.

\begin{widetext}

\appendix*

\section{ Heavy quark propagator and different correlation functions}

\renewcommand{\theequation}{\Alph{section}.\arabic{equation}} \label{sec:App}


In the present paper, for the heavy quark propagator $S_{Q}^{ab}(x)$
($Q=c,\ b$), we employ the following expression
\begin{eqnarray}
&&S_{Q}^{ab}(x)=i\int \frac{d^{4}k}{(2\pi )^{4}}e^{-ikx}\Bigg \{\frac{\delta
_{ab}\left( {\slashed k}+m_{Q}\right) }{k^{2}-m_{Q}^{2}}-\frac{%
g_{s}G_{ab}^{\alpha \beta }}{4}\frac{\sigma _{\alpha \beta }\left( {\slashed %
k}+m_{Q}\right) +\left( {\slashed k}+m_{Q}\right) \sigma _{\alpha \beta }}{%
(k^{2}-m_{Q}^{2})^{2}}  \notag \\
&&+\frac{g_{s}^{2}G^{2}}{12}\delta _{ab}m_{Q}\frac{k^{2}+m_{Q}{\slashed k}}{%
(k^{2}-m_{Q}^{2})^{4}}+\cdots \Bigg \}.
\end{eqnarray}%
Here, we have introduced the notations
\begin{equation}
G_{ab}^{\alpha \beta }\equiv G_{A}^{\alpha \beta }\lambda _{ab}^{A}/2,\ \
G^{2}=G_{\alpha \beta }^{A}G_{A}^{\alpha \beta },\
\end{equation}%
with $G_{A}^{\alpha \beta }$ being the gluon field-strength tensor, and $%
\lambda ^{A}$--Gell-Mann matrices. The index $A$ varies in the range $%
1-8$.

This Appendix also contains expressions of correlation functions, which are
employed to calculate some of the strong couplings. In the case of the decay
$X_{\mathrm{4c}}^{\ast }\rightarrow \chi _{c0}\chi _{c0}$ the correlation
function $\Pi ^{\mathrm{OPE}}(p,p^{\prime })$ is given by the formula
\begin{eqnarray}
&&\Pi ^{\mathrm{OPE}}(p,p^{\prime })=2i^{2}\int
d^{4}xd^{4}ye^{-ipx}e^{ip^{\prime }y}\left\{ \mathrm{Tr}\left[
S_{c}^{ia}(y-x)\gamma _{\mu }\widetilde{S}_{c}^{jb}(-x)\widetilde{S}%
_{c}^{bj}(x)\gamma ^{\mu }S_{c}^{ai}(x-y)\right] \right.  \notag \\
&&\left. -\mathrm{Tr}\left[ S_{c}^{ia}(y-x)\gamma _{\mu }\widetilde{S}%
_{c}^{jb}(-x)\widetilde{S}_{c}^{aj}(x)\gamma ^{\mu }S_{c}^{bi}(x-y)\right]
\right\} ,  \label{eq:ACF1}
\end{eqnarray}%
where $\widetilde{S}_{c}(x)=CS_{c}(x)C,$ and $C$ is the charge conjugation
operator. The fit function $\mathcal{G}_{6}^{\ast }(Q^{2})$ used to find the
strong coupling $g_{6}^{\ast }$ is fixed by the parameters $\mathcal{G}%
_{6}^{0\ast }=0.51~\mathrm{GeV}^{-1}$, $c_{6}^{1\ast }=3.11$, and $%
c_{6}^{2\ast }=-3.57$.

For the decay $X_{\mathrm{4c}}^{\ast }\to \chi _{c1}\chi _{c1}$ the function
$\Pi _{\mu \nu }^{\mathrm{OPE}}(p,p^{\prime })$ has the following form:

\begin{eqnarray}
&&\Pi _{\mu \nu }^{\mathrm{OPE}}(p,p^{\prime })=2i^{2}\int
d^{4}xd^{4}ye^{-ipx}e^{ip^{\prime }y}\left\{ \mathrm{Tr}\left[ \gamma _{\mu
}\gamma _{5}S_{c}^{ia}(y-x)\gamma _{\alpha }\widetilde{S}_{c}^{jb}(-x)\gamma
_{5}\gamma _{\nu }\widetilde{S}_{c}^{aj}(x)\gamma ^{\alpha }S_{c}^{bi}(x-y)%
\right] \right.  \notag \\
&&\left. -\mathrm{Tr}\left[ \gamma _{\mu }\gamma _{5}S_{c}^{ia}(y-x)\gamma
_{\alpha }\widetilde{S}_{c}^{jb}(-x)\gamma _{5}\gamma _{\nu }\widetilde{S}%
_{c}^{bj}(x)\gamma ^{\alpha }S_{c}^{ai}(x-y)\right] \right\} .
\label{eq:ACF2}
\end{eqnarray}%
In this case, the function $\mathcal{G}_{7}^{\ast }(Q^{2})$ has the
parameters: $\mathcal{G}_{7}^{0\ast }=0.74~\mathrm{GeV}^{-1}$, $c_{7}^{1\ast
}=2.48$, and $c_{7}^{2\ast }=-3.01$.

The correlation functions for the decays of the hadronic molecule $\mathcal{M%
}$ and functions to calculate the relevant strong couplings:

Decay $\mathcal{M}\to \eta _{c}\chi _{c1}$%
\begin{eqnarray}
&&\Pi _{\mu }^{\mathrm{OPE}}(p,p^{\prime })=2i^{3}\int
d^{4}xd^{4}ye^{-ipx}e^{ip^{\prime }y}\left\{ \mathrm{Tr}\left[ \gamma _{\mu
}\gamma _{5}S_{c}^{ia}(y-x)\gamma _{\alpha }\gamma _{5}S_{c}^{ai}(x-y)\right]
\mathrm{Tr}\left[ \gamma _{5}S_{c}^{jb}(-x)\gamma ^{\alpha }\gamma
_{5}S_{c}^{bj}(x)\right] \right.  \notag \\
&&\left. -\mathrm{Tr}\left[ \gamma _{\mu }\gamma _{5}S_{c}^{ia}(y-x)\gamma
_{\alpha }\gamma _{5}S_{c}^{aj}(x)\gamma _{5}S_{c}^{jb}(-x)\gamma ^{\alpha
}\gamma _{5}S_{c}^{bi}(x-y)\right] \right\} ,  \label{eq:A4}
\end{eqnarray}%
and the fit function $\mathcal{G}_{10}(Q^{2})$ for $G_{5}(Q^{2})$: $\mathcal{%
G}_{10}^{0}=3.85$, $c_{10}^{1}=3.51$, and $c_{10}^{2}=-4.56$.

Decay $\mathcal{M}\to \chi _{c0}\chi _{c0}$%
\begin{equation}
\Pi ^{\mathrm{OPE}}(p,p^{\prime })=-2i^{2}\int
d^{4}xd^{4}ye^{-ipx}e^{ip^{\prime }y}\mathrm{Tr}\left[ S_{c}^{ia}(y-x)\gamma
_{\alpha }\gamma _{5}S_{c}^{aj}(x)S_{c}^{jb}(-x)\gamma ^{\alpha }\gamma
_{5}S_{c}^{bi}(x-y)\right] ,  \label{eq:A5}
\end{equation}%
and $G_{6}(Q^{2})$: $\mathcal{G}_{11}^{0}=0.55~\mathrm{GeV}^{-1}$, $%
c_{11}^{1}=3.06$, and $c_{11}^{2}=-3.46$.

Decay $\mathcal{M}\to \chi _{c1}\chi _{c1}$%
\begin{eqnarray}
\widehat{\Pi }_{\mu \nu }^{\mathrm{OPE}}(p,p^{\prime }) &=&2i^{2}\int
d^{4}xd^{4}ye^{-ipx}e^{ip^{\prime }y}\left\{ \mathrm{Tr}\left[ \gamma _{\mu
}\gamma _{5}S_{c}^{ia}(y-x)\gamma _{\alpha }\gamma _{5}S_{c}^{ai}(x-y)\right]
\mathrm{Tr}\left[ \gamma _{\nu }\gamma _{5}S_{c}^{jb}(-x)\gamma ^{\alpha
}\gamma _{5}S_{c}^{bj}(x)\right] \right.  \notag \\
&&\left. -\mathrm{Tr}\left[ \gamma _{\mu }\gamma _{5}S_{c}^{ia}(y-x)\gamma
_{\alpha }\gamma _{5}S_{c}^{aj}(x)\gamma _{\nu }\gamma
_{5}S_{c}^{jb}(-x)\gamma ^{\alpha }\gamma _{5}S_{c}^{bi}(x-y)\right]
\right\} ,  \label{eq:A6}
\end{eqnarray}%
and the parameters $\mathcal{G}_{12}^{0}=0.86~\mathrm{GeV}^{-1}$, $%
c_{12}^{1}=2.41$, and $c_{12}^{2}=-2.89$ to compute $G_{7}(Q^{2})$.

\end{widetext}

\end{document}